# Elastic and electronic properties of hexagonal rhenium sub-nitrides Re$_3$N and Re$_2$N in comparison with *hcp*-Re and wurtzite-like rhenium mononitride ReN

V. V. Bannikov, I. R. Shein, * A. L. Ivanovskii

Institute of Solid State Chemistry, Ural Branch of the Russian Academy of Sciences, 620990, Ekaterinburg, Russia

**A B S T R A C T**

Very recently, two new hexagonal rhenium sub-nitrides Re$_3$N and Re$_2$N, which belong to a rather rare group of known metal-rich (M/N > 1) nitrides of heavy 4$d$,5$d$ metals, have been successfully synthesized, and their potential technological applications as ultra-incompressible materials have been proposed. In this work we present a detailed *ab initio* study of novel rhenium sub-nitrides in comparison with *hcp*-Re and wurtzite-like rhenium mono-nitride ReN, with the purpose to evaluate the trends of the elastic, electronic properties and chemical bonding in the series of these hexagonal systems as a function of the Re/N stoichiometry: Re→ Re$_3$N → Re$_2$N → ReN.

* Corresponding author.
*E-mail address:* shein@ihim.uran.ru (I.R. Shein).

# 1. Introduction.

Intensive experimental and theoretical efforts focused today on the design of new compounds of heavy 4*d*- (Tc-Pd) or 5*d*- (Re-Pt) metals (*M*) with light *sp* elements - boron, carbon, and nitrogen, are motivated by the search of new ultra-incompressible and superhard materials for various technological applications, reviews [1-3]. Recently, a series of clear experimental evidences of enhanced mechanical properties of such materials as $ReB_2$ [4,5], $OsB_2$ [6,7], $WB_4$-like tetraborides, and a set of mono- and dinitrides and carbides of some heavy 4*d* and 5*d* metals [2,3,8], has been presented. Simultaneously the theoretical backgrounds of design principles for such materials are actively discussed [1-3,9,10].

Let us note that the stoichiometry of the above materials is mostly $(B,C,N)/M = 1 \div 4$ [1-8], whereas only in few works [11-19] metal-rich systems ($(B,C,N)/M < 1$) have been discussed.

In this context, the recent synthesis [20] of two novel rhenium sub-nitrides, namely hexagonal $Re_3N$ and $Re_2N$, seems to be highly interesting for further insight into the nature of such metal-rich systems.

Motivated by these circumstances, in this work we present a detailed *ab initio* study of the above novel rhenium sub-nitrides in comparison with *hcp*-Re and wurtzite-like rhenium mono-nitride ReN with the purpose to found out the trends of the elastic and electronic properties in the series of these hexagonal systems as a function of the Re/N stoichiometry: Re→ $Re_3N$ → $Re_2N$ → ReN.

# 2. Models and computational aspects.

The recently discovered rhenium sub-nitrides $Re_3N$ and $Re_2N$ [20] adopt the hexagonal structures with space groups *P-6m2* and *P6_3/mmc*, respectively. The atomic positions are [20]: for $Re_3N$ – Re: (⅔; ⅓; ½) and (⅓; ⅔; *z*) with *z*=0.198, N: (⅔; ⅓; 0); for $Re_2N$ – Re: (⅓; ⅔; *z*) with *z*=0.106, and N: (⅓; ⅔; ¾).

In addition, metallic Re and ReN have been investigated. It is well known that metallic Re adopts a *hcp*-structure. For the rhenium mononitride among the proposed series of polymorphs, we have chosen for our analysis the wurtzite-like ReN, which was predicted [21] to be the most stable. Note that in Ref. [16] it was reported that the zinc blende structure is energetically more preferable for ReN, but this polymorph appears to be mechanically unstable [21].

For the calculations of the elastic parameters of hexagonal Re, $Re_3N$, $Re_2N$ and ReN, we employed the Vienna *ab initio* simulation package (VASP) in projector augmented waves (PAW) formalism [22,23]. Exchange and correlation were described by a non-local correction for LDA in the form of GGA [24]. The kinetic energy cutoff of 500 eV and k-mesh of 16×16×6 were used. The geometry optimization was performed with the force cutoff of 2 meV/Å.

Furthermore, the electronic properties of the above materials were examined by means of the full-potential method with mixed basis APW+lo (FLAPW-GGA) implemented in the WIEN2k suite of programs [25]. The plane-wave expansion with $R_{MT} \times K_{MAX}$ was equal to 7.0, and $k$ sampling with 1500 points in the full Brillouin zone was used. The MT sphere radii were chosen to be 1.5 bohr for nitrogen and 2.2 bohr for rhenium. The Blöchl's modified tetrahedron method [26] was employed to calculate the densities of states (DOS). The calculations were considered to be converged when the difference in the total energy did not exceed 0.01 mRy, 0.001$e$ in electronic charge and 1 mRy/bohr in forces as calculated at consecutive steps.

These two DFT-based codes are complementary and allow us to perform a complete investigation of the declared properties of the above materials.

## 3. Results and discussion.

*3.1. Lattice parameters and density.*
At the first stage the equilibrium lattice constants ($a$ and $c$) and cell volumes ($V$) for hexagonal rhenium, $Re_3N$, $Re_2N$, and $ReN$ were found. The calculated values are presented in Table 1 and are in reasonable agreement with the available experiments and earlier calculations. The obtained equilibrium cell volumes were used for the estimation of theoretical density ($\rho^{theor}$) of the investigated systems; the calculated data (Table 1) show that $\rho^{theor}$ decreases with reduction of the Re content, *i.e.* in the sequence $Re > Re_3N > Re_2N > ReN$.

*3.2. Elastic properties.*
Let us discuss the trends in the elastic parameters for the series of hexagonal rhenium-nitrogen systems Re, $Re_3N$, $Re_2N$, and ReN. The calculated values of five independent elastic constants for hexagonal crystals ($C_{11}$, $C_{12}$, $C_{13}$, $C_{33}$, and $C_{44}$) are given in Table 2. First of all, $C_{ij}$ constants for all the systems are positive and satisfy the generalized criteria [33] for mechanically stable crystals: $C_{44} > 0$, $C_{11} > |C_{12}|$, and $(C_{11} + 2C_{12})C_{33} > 2C_{13}^2$. Further, the calculated elastic constants allow us to obtain the bulk $B$ and shear $G$ moduli of these materials. Usually, for such estimations two main approximations are employed, namely, the Voigt (V) [34] and Reuss (R) [35] schemes. Thus, in terms of the Voigt approximation, the $B$ and $G$ values are obtained as:
$B_V = (1/9) \{2(C_{11} + C_{12}) + 4C_{13} + C_{33}\}$,
$G_V = (1/30) \{C_{11} + C_{12} + 2C_{33} - 4C_{13} + 12C_{44} + 12C_{66}\}$,
and in terms of the Reuss approximation as:
$B_R = \{(C_{11} + C_{12})C_{33} - 2C_{12}^2\}/(C_{11} + C_{12} + 2C_{33} - 4C_{13})$,
$G_R = (5/2) \{[(C_{11} + C_{12})C_{33} - 2C_{12}^2] C_{55}C_{66}\}/\{3B_V C_{55}C_{66} + [(C_{11} + C_{12})C_{33} - 2C_{12}^2]^2(C_{55} + C_{66})\}$.
The synthesized $Re_3N$ and $Re_2N$ sub-nitrides are prepared and investigated [20] as polycrystalline species, *i.e.* in the form of aggregated mixtures of microcrystalline grains with random orientation. For sure estimations of the elastic parameters for the polycrystalline materials, the Voigt-Reuss-Hill (VRH) approximation [36] is widely used, where the effective moduli for isotropic polycrystals are expressed as the arithmetic mean of the two above mentioned (Voigt and Reuss) limits: $B = 1/2(B_V + B_R)$ and $G = 1/2(G_V + G_R)$, and averaged compressibility is $\beta = 1/B$.

Further, the calculated isotropic bulk moduli $B$ and shear moduli $G$ allow us to obtain the Young's moduli $Y$ and the Poisson's ratio $v$ as:
$Y = 9BG/(3B + G)$,
$v = (3B - 2G)/\{2(3B + G)\}$.

The above elastic parameters presented in Table 3 allow us to make the following conclusions:

(i). For all of the examined systems, $B > G$; this implies that the parameter limiting the mechanical stability of these materials is the shear modulus. In turn, the bulk modulus $B$ (compressibility $\beta$) in the series of these hexagonal systems as a function of the Re/N stoichiometry: Re→ $Re_3N$ → $Re_2N$ → ReN has the maximal (minimal) value for the sub-nitrides $Re_3N$ and $Re_2N$. These phases adopt the bulk moduli $B \geq 400$ GPa - close to those for the most incompressible binary $d$ metal carbides and nitrides found to date [20]. The enhancement of the bulk moduli (incompressibility) for $Re_3N$ and $Re_2N$ in comparison with metallic rhenium and ReN may be explained in a simple way taking into account the known direct correlation between $B$ and the effective density of valence electrons, see [34]. According to our estimations, the average concentration of valence electrons (in $e/Å^3$) for these sub-nitrides is maximal: 0.444 (ReN) < 0.470 (Re) < 0.482 ($Re_3N$) < 0.487 ($Re_2N$). Besides, the shear modulus $G$ and the Young modulus $Y$ vary in non-monotonic manner with maxima for the sub-nitrides $Re_3N$ and $Re_2N$, see Table 3.

(ii). *Intrinsic* hardness together with compressibility belongs to the most important mechanical characteristics of materials. However, hardness is a macroscopic concept, which is characterized experimentally by indentation, and thus depends strongly on plastic deformation – unlike compressibility related to elastic deformation, see [10,35]. Even though these types of deformations (plastic and elastic) are fundamentally different, the values of the bulk moduli $B$ (which measures the resistance to volume change with invariable proportions) and the shear moduli $G$ (which measures the resistance to shear deformation) are often used [36-38] as preliminary hardness predictors. In our case, both $B$ and $G$ moduli adopt the maximal values for $Re_3N$ and $Re_2N$, thus these materials may be expected to exhibit enhanced hardness.

(iii). Another important mechanical characteristic of materials is their brittle/ductile behavior, which is closely related, in particular, to their reversible compressive deformation and fracture ability. Here, the widely used malleability measures are the Pugh's indicator ($G/B$ ratio) [39] and the so-called machinability index $\mu_M = B/C_{44}$ [40]. As is known empirically, if $G/B < 0.5$, a material behaves in a ductile manner, and *vice versa*, if $G/B > 0.5$, a material demonstrates brittleness. Our data reveal that for all of the examined systems $G/B \sim 0.5$, *i.e.* Re nitrides are close to the brittle/ductile border, but for the sub-nitrides some growth of their brittleness may be expected, meanwhile both metallic rhenium and ReN behave rather in a ductile manner, see Table 3. On the other hand, the machinablity index $\mu_M$ is useful to compare the relative malleability in this series of materials. According to our estimations, both the *hcp*-Re and novel sub-nitrides are characterized by comparable values of $\mu_M$ index (~ 2.0-2.2), so it is reasonable to expect that the malleability of $Re_3N$ and $Re_2N$ is similar to that of metallic rhenium, meanwhile for ReN the value of $\mu_M$ is about 1.5 times greater, which qualitatively correlates with *G/B* ratio values indicating that ReN should be the most ductile material in the series. An additional indicator of brittle/ductile behavior follows from the Poisson's ratio ν: its values for brittle covalent materials are small (~0.1), whereas for ductile metallic materials ν is typically ~0.33 [36]. The calculated Poisson's ratio varies in a rather narrow interval 0.27-0.31 (Table 3), but for the sub-nitrides these values are minimal. This means a growth of covalence and, therefore, the increased brittleness.

(iv). The elastic anisotropy of crystals is an important parameter for engineering science since it correlates with the possibility of microcracks appearance in materials [41,42]. There are different ways to represent the elastic anisotropy of crystals, for example, using the calculated $C_{ij}$ constants. So, the shear anisotropic factors can be obtained as a measure of the degree of anisotropy in the bonding between atoms in different planes [43,44]. In particular, for the {100} shear planes between ‹011› and ‹010› directions the shear anisotropic factor *A* is defined as: $A = 4C_{44}/(C_{11} + C_{33} - 2C_{13})$. For crystals with isotropic elastic properties $A = 1$, while values smaller or greater than unity measure the degree of elastic anisotropy. Another way implies the use of the parameter *f* which is defined for hexagonal crystals as: $f = (C_{11} + C_{12} - 2C_{13})/(C_{33} - C_{13})$. It describes the ratio between linear compressibility coefficients of hexagonal crystals; and it is assumed that the value $f = 1$ corresponds to isotropic compressibility, while the deviation from the unity is a measure of anisotropy for linear compressibility along the *c* and *a* axes [45]. Finally, so-called universal anisotropy index defined as: $A^U = 5G_V/G_R + B_V/B_R - 6$ was proposed recently [46]. For isotropic crystals $A^U = 0$; deviations of $A^U$ from zero define the extent of crystal anisotropy.

In our case, the calculated indexes *A*, *f*, and $A^U$ listed in Table 3 demonstrate that the minimal anisotropy among the examined materials should be expected for $Re_3N$ and $Re_2N$, *i.e.* the elastic properties of novel rhenium sub-nitrides should be more isotropic than those for metallic rhenium and wurtzite-like ReN.

*3.3. Electronic structure and inter-atomic bonding.*

The total and site- and *l*-projected densities of states (DOSs) for $Re_2N$ and $Re_3N$ as well as for Re and ReN are presented in Fig. 1. It is seen that as compared with Re, the spectra of rhenium nitrides are more complicated owing to the formation of new bands originating from nitrogen *s,p* states accompanied by an appreciable redistribution of near-Fermi Re *5d* states. According to our calculations, the sub-nitrides $Re_2N$ and $Re_3N$ are non-magnetic and should exhibit metallic conductivity.

The near-Fermi bands of $Re_2N$ and $Re_3N$ are composed mainly of Re 5*d* states, while the role of N 2*p* states is relatively small. In turn these N 2*p* states, giving the considerable contributions to the bottom of the valence band, are hybridized with Re 5*d* states and are responsible for the formation of covalent Re-N bonds.

As going from *hcp*-Re to ReN mononitride, the DOSs values at the Fermi level ($N(E_F)$, per Re atom) vary non-monotonously, slightly decreasing for $Re_2N$ and $Re_3N$ and then increasing for ReN. Thus, there is no simple trend in the electronic properties depending on the N/Re stoichiometry. Probably, the strongest influence here is exerted by the structural factor, when $Re_2N$ and $Re_3N$ adopt the lattices with alternating rhenium and nitrogen layers [20], rarely observed for transition metal nitrides.

Finally, let us discuss the bonding picture in $Re_2N$ and $Re_3N$ sub-nitrides. Our analysis reveals that bonding in these species is of a complex character and may be described as a mixture of metallic, ionic, and covalent contributions. So, the delocalized metallic-like Re-Re interactions appear owing to near-Fermi Re 5*d* states. The character of the Re-N covalent bonding (owing to the above hybridization of N 2*p* - Re 5*d* states) may be well illustrated using the charge density maps, Fig. 2. Besides, the formation of the directional bonds between Re atoms is distinctly visible. It may be assumed that these directional Re-N and Re-Re bonds together with the above enhanced effective density of valence electrons are favorable factors for the growth of elastic moduli and incompressibility of $Re_2N$ and $Re_3N$.

To describe the ionic bonding for these materials, we carried out a Bader [48] analysis, and the total charges of the atoms in crystal (the so-called Bader charges $Q^B$) as well as the effective charges estimated as: $\Delta Q^{eff} = Q^B - Q^n$, where $Q^n$ is the total charge of the corresponding neutral atom, are presented in Table 4. These results show that a considerable Re → N charge transfer takes place for all the nitrides, which is however much smaller than predicted within the simple ionic model. For example, the calculated values of $\Delta Q^{eff}$ for nitrogen atoms in Re nitrides vary over the range 1.1 *e* - 1.3 *e versus* 3 *e* as assumed in the purely ionic model. The positive charges $\Delta Q^{eff}$ of Re atoms in $Re_2N$ and $Re_3N$ are considerably smaller (~0.1 - 0.6 *e*), increasing to ~1.1 *e* as going to the rhenium mononitride. Let us note that a considerable difference in $\Delta Q^{eff}$ is obtained for non-equivalent Re atoms in $Re_3N$, where higher positive charges (~ 0.6 *e*) are adopted by rhenium atoms nearest to nitrogen atoms.

## 4. Conclusions.

In summary, we shall outline the main results of our study. Firstly, we found that in the series of hexagonal systems Re→ $Re_3N$ → $Re_2N$ → ReN their bulk moduli *B* (compressibility *β*) have maximal (minimal) values for sub-nitrides $Re_3N$ and $Re_2N$ – owing to the formation of directional Re-N and Re-Re bonds and to enhanced effective density of valence electrons. As the shear moduli *G* adopt the maximal values also for Re$_3$N and Re$_2$N, these materials may be expected to exhibit enhanced hardness. Further, these sub-nitrides are close to the brittle/ductile border, and their elastic anisotropy will be smaller than for *hcp*-Re and wurtzite-like ReN.

According to our calculations, the sub-nitrides Re$_2$N and Re$_3$N are non-magnetic and should exhibit metallic conductivity. Their near-Fermi bands are composed mainly of Re *5d* states. Our analysis reveals that bonding in these species is of a complex character and may be described as a mixture of metallic, ionic, and covalent contributions, where the delocalized metallic-like Re-Re interactions appear owing to near-Fermi Re *5d* states, and the ionic bonding is due to Re → N charge transfer. Our estimations show that this charge transfer is much smaller than predicted within the simple ionic model, and the effective ionic charges for non-equivalent Re atoms in Re$_3$N differ considerably.

**Table 1.** Calculated lattice constants ($a$, $c$, in Å), cell volume ($V$, in Å$^3$ per formula unit) and density ($\rho$, in g/cm$^3$) of *hcp*-Re and hexagonal rhenium nitrides in comparison with available experimental and theoretical data.

| system | Re | Re$_3$N | Re$_2$N | ReN |
|---|---|---|---|---|
| $a$ | 2.766 * (2.76 [a]; 2.776 [b]; 2.758 [c]) | 2.828 (2.78 [e]; 2.8065 [f]) | 2.857 (2.83 [e]; 2.837 [f]) | 2.777 (2.750 [g]) |
| $c$ | 4.492 (4.46 [a]; 4.472 [b]; 4.447 [c]) | 7.185 (7.152 [e]; 7.112 [f]) | 9.877 (9.88 [e]; 9.799 [f]) | 6.739 (6.641 [g]) |
| $c/a$ | 1.624 (1.615 [d]) | 2.541 | 3.457 | 2.427 |
| $V$ | 14.88 | 49.76 | 34.91 | 22.50 |
| $\rho$ | 20.78 (21.03 [a]) | 19.11 | 18.38 | 14.78 |

\* our VASP data
[a] Ref. [27] – experiment
[b] Ref. [16] – calculated, CASTEP-GGA
[c] Ref. [28] – estimated at 4.2 K
[d] Ref. [29] – calculated, LAPW-LDA, -GGA
[e, f] Ref. [20] – experiment and calculated (CASTEP-GGA), respectively
[g] Ref. [21] – calculated, CASTEP-GGA

**Table 2.** Calculated elastic constants ($C_{ij}$, in GPa) of *hcp*-Re and hexagonal rhenium nitrides in comparison with available experimental and theoretical data.

| | Re | Re$_3$N | Re$_2$N | ReN |
|---|---|---|---|---|
| $C_{11}$ | 591 (612.5 [a]; 617.7 [b]; 837.3 [c]; 605 [d]; 640 [e]) | 657 | 662 | 570 (562 [f]) |
| $C_{33}$ | 793 (682.7 [a]; 682.8 [b]; 894.6 [c]; 650 [d]; 695 [e]) | 845 | 870 | 794 (777 [f]) |
| $C_{44}$ | 162 (162.5 [a]; 160.5 [b]; 222.5 [c]; 175 [d]; 170 [e]) | 198 | 192 | 110 (106 [f]) |
| $C_{12}$ | 361 (270 [a]; 274.9 [b]; 293.3 [c]; 235 [d]; 280 [e]) | 248 | 237 | 252 (233 [f]) |
| $C_{13}$ | 203 (206 [a]; 205.6 [b]; 216.8 [c]; 195 [d]; 220 [e]) | 244 | 258 | 202 (206 [f]) |

[a] Ref. [30] – experiment
[b] Ref. [31] – experiment
[c] Ref. [32] – calculated, FP-LMTO
[d, e] Ref. [29] – calculated, FP-LAPW, LDA/GGA, respectively
[f] Ref. [21] – calculated, CASTEP-GGA

**Table 3**. Calculated bulk ($B_V$, $B_R$, and $B$) and shear moduli ($G_V$, $G_R$, and $G$) in Voigt, Reuss and Voigt-Reuss-Hill approximations (in GPa), compressibility ($\beta$, in GPa$^{-1}$), Young modulus ($Y$, in GPa), Poisson`s ratio ($\nu$), Pugh's indicator ($G/B$ ratio), machinability index ($\mu_M$) and indexes of elastic anisotropy ($A$, $f$, and $A^U$) for *hcp*-Re and hexagonal rhenium nitrides in comparison with available experimental and theoretical data.

| system | Re | Re$_3$N | Re$_2$N | ReN |
|---|---|---|---|---|
| $B_V$ | 389.9 | 403.4 | 411.1 | 360.7 |
| $B_R$ | 286.4 | 396.4 | 416.8 | 328.1 |
| $B_{VRH}$ | 338.1 (365.2 [a]; 372 [b]; 447.3 [c]; 382 [d]; 344 [e]) | 399.9 (395 [f]; 413 [g]) | 413.9 (401 [f]; 415 [g]) | 344.40 (349 [h]) |
| $\beta$ | 0.002957 | 0.002501 | 0.002416 | 0.002904 |
| $G_V$ | 168.3 | 215.0 | 215.4 | 161.0 |
| $G_R$ | 145.1 | 211.4 | 212.7 | 143.4 |
| $G_{VRH}$ | 156.7 (178.9 [a]) | 213.2 | 214.0 | 152.2 (151 [h]) |
| $Y$ | 40729 | 543.0 | 547.7 | 397.9 (396 [h]) |
| $\nu$ | 0.2993 (0.2894 [b]) | 0.2737 | 0.2795 | 0.3074 (0.31 [h]) |
| $G/B$ | 0.46 | 0.53 | 0.52 | 0.44 |
| $\mu_M$ | 2.09 | 2.02 | 2.16 | 3.13 |
| $A$ | 0.66 | 0.78 | 0.76 | 0.46 |
| $f$ | 0.93 | 0.69 | 0.63 | 0.71 |
| $A^U$ | 0.44 | 0.11 | 0.03 | 0.73 |

[a] Ref. [31] – experiment
[b] Ref. [27] – experiment
[c] Ref. [32] – calculated, FP-LMTO
[d, e] Ref. [29] – calculated, FP-LAPW, LDA/GGA, respectively
[f, g] Ref. [20] – experiment and calculated (CASTEP-GGA), respectively
[h] Ref. [21] – calculated, CASTEP-GGA

**Table 4.** Calculated total densities of states at the Fermi level ($N(E_F)$, in states/eV per formula unit), so-called Bader charges ($Q^B$, in $e$), and effective atomic charges ($\Delta Q^{eff}$ in $e$) for *hcp*-Re and hexagonal rhenium nitrides.

| | Re | Re$_3$N | Re$_2$N | ReN |
|---|---|---|---|---|
| $N(E_F)$ | 0.731 (0.740[a]) | 2.010 | 1.020 | 1.036 |
| $Q^B$(Re) | - | 74.90/74.42 [b] | 74.35 | 73.88 |
| $\Delta Q^{eff}$(Re) | - | + 0.10/+ 0.58 [b] | + 0.65 | +1.12 |
| $Q^B$(N) | - | 8.27 | 8.29 | 8.12 |
| $\Delta Q^{eff}$(N) | - | - 1.27 | - 1.29 | - 1.12 |

[a] Ref. [28] – calculated, relativistic APW
[b] for non-equivalent Re atoms in Re$_3$N, placed in positions: (⅔; ⅓; ½)/(⅓; ⅔; $z$)

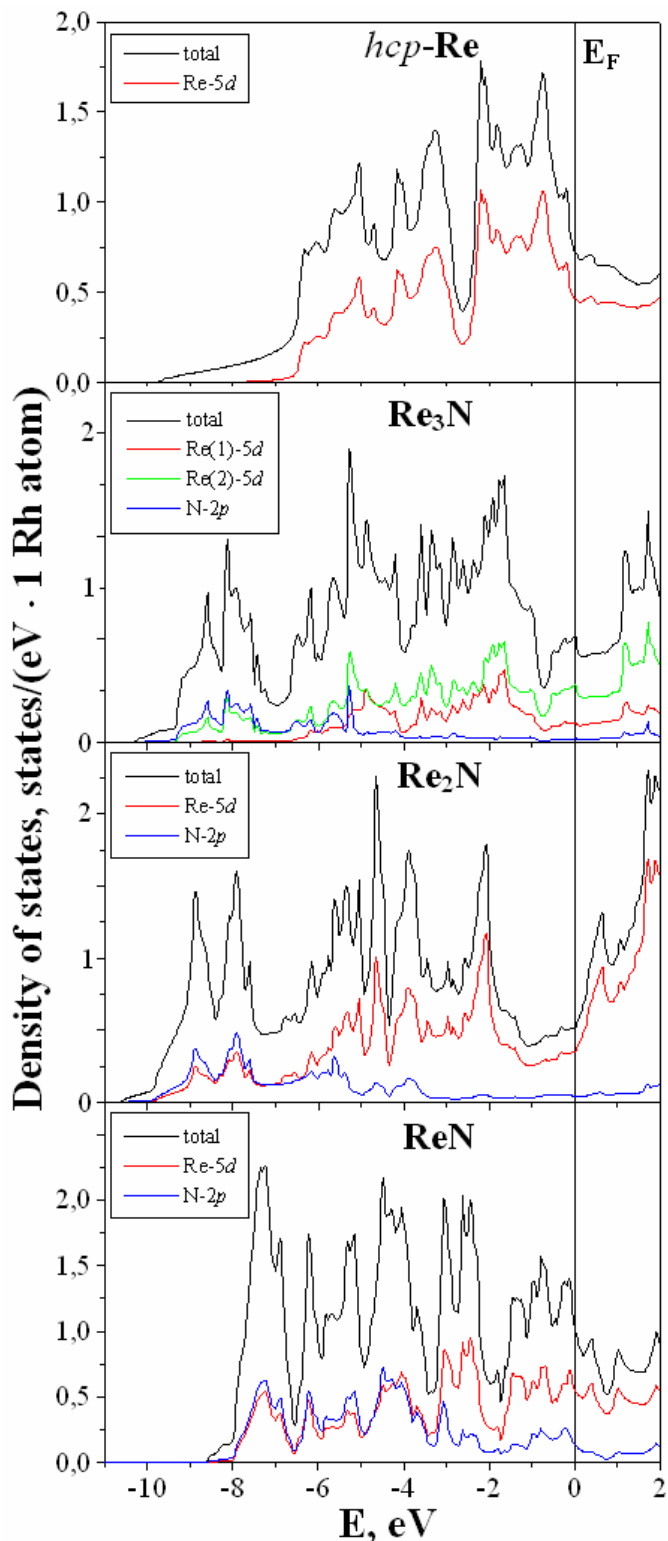

**Fig.1.** Total and partial densities of states of *hcp*-Re and hexagonal rhenium nitrides Re$_3$N, Re$_2$N, and ReN (related to one Re atom) as obtained within FLAPW-GGA calculations. For Re$_3$N, Re(1,2) denote the states of Re in (⅔; ⅓; ½) and (⅓; ⅔; z) positions, respectively.

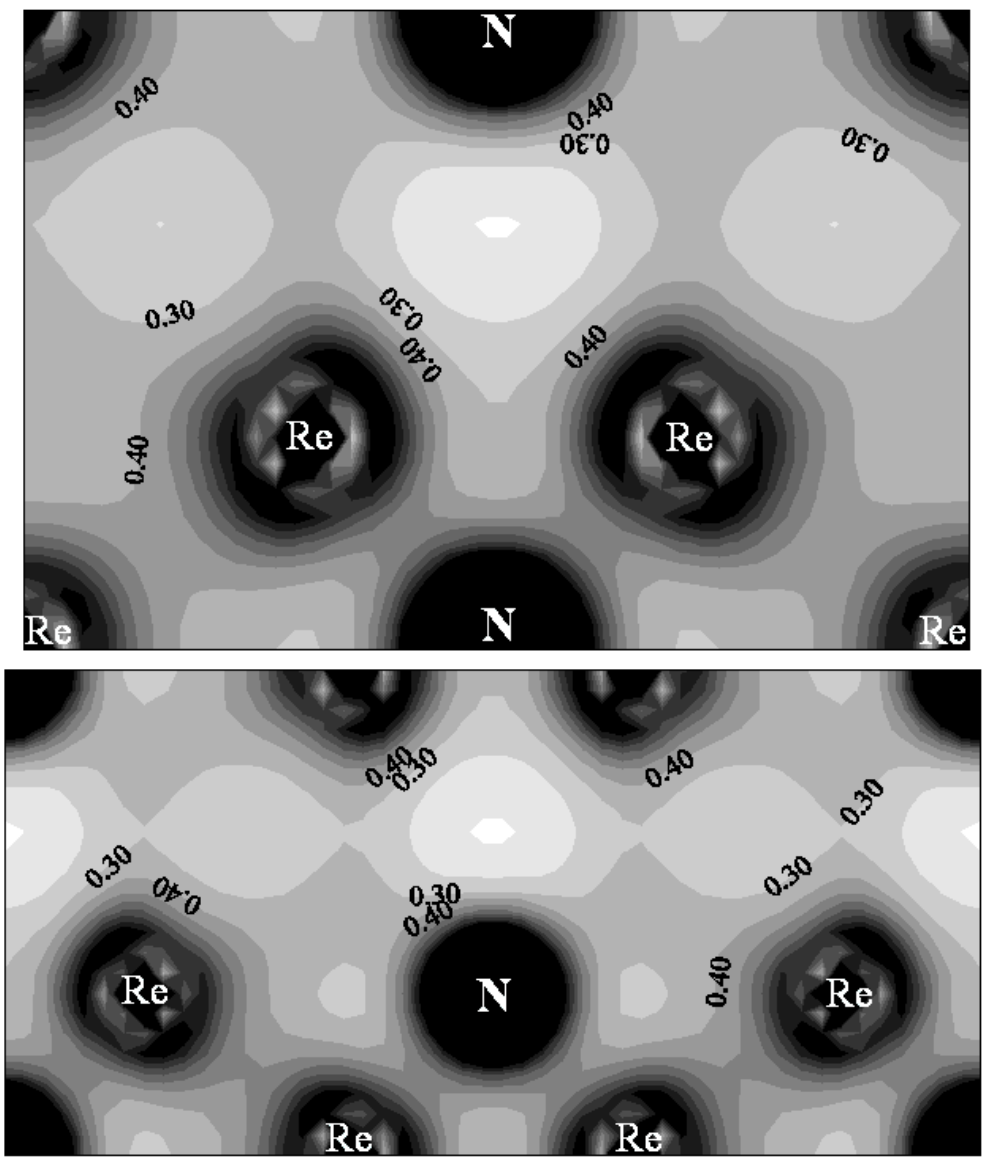

**Fig.2.** The charge density maps for Re-N planes parallel to *z* axis in Re$_3$N (*upper map*) and Re$_2$N (*lower map*).